\documentclass[a4paper,11pt]{article}
\usepackage{graphicx,amssymb,amsmath,amscd,graphicx,epic,array,wrapfig}
\newcommand{\DIS}{\displaystyle}

\def\C{{\mathbb C}}
\def\Z{{\mathbb Z}}
\def\R{{\mathbb R}}

\def\P{{\mathbb P}}
\def\T{{\mathbb T}}
\def\F{{\mathbb F}}
\def\H{{\mathbb H}}

\def\bx{\text{\mathversion{bold}{$x$}}}

\newtheorem{theorem}{Theorem}

\begin{document}
\title{\textbf{A tropical analogue of the Hessian group}}
\author{\sc NOBE Atsushi\\ Department of Mathematics, Faculty of Education, Chiba University,\\
1-33 Yayoi-cho Inage-ku, Chiba 263-8522, Japan}
\date{}
\maketitle

\begin{abstract}
We investigate a tropical analogue of the Hessian group $G_{216}$, the group of linear automorphisms acting on the Hesse pencil. 
Through the procedure of ultradiscretization, the group law on the Hesse pencil reduces to that on the tropical Hesse pencil.
We then show that the dihedral group $\mathcal{D}_3$ of degree three is the group of linear automorphisms acting on the tropical Hesse pencil.
\end{abstract}

\begin{wrapfigure}[15]{r}{7cm}
\centering
{
\includegraphics[scale=.7]{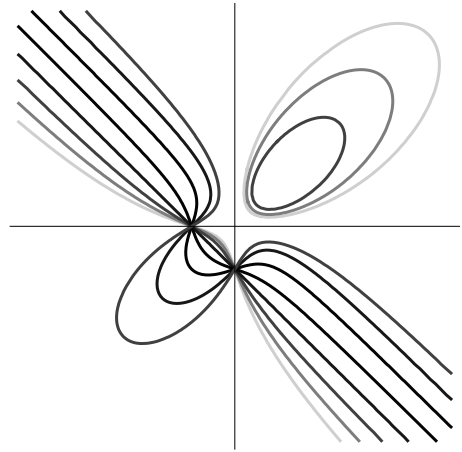}
}
\caption{Several members of the Hesse pencil. 
}
\label{fig:HesseCurve}
\end{wrapfigure}

The Hessian group $G_{216}\simeq \Gamma\rtimes SL(2,\F_3)$ is a subgroup of $PGL(3,\C)$, the group of linear transformations on the projective plane $\P^2(\C)$, where $\Gamma=\left(\Z/3\Z\right)^2$ and $SL(2,\F_3)$ is the special linear group over the finite field $\F_3$ of characteristic three.
The Hessian group is generated by the following four linear transformations
\begin{align*}
g_1
=
\left(\begin{matrix}
0&1&0\\
0&0&1\\
1&0&0\\
\end{matrix}\right)
\qquad
g_2
=
\left(\begin{matrix}
1&0&0\\
0&\zeta_3&0\\
0&0&\zeta_3^2\\
\end{matrix}\right)
\qquad
g_3
=
\left(\begin{matrix}
1&1&1\\
1&\zeta_3&\zeta_3^2\\
1&\zeta_3^2&\zeta_3\\
\end{matrix}\right)
\qquad
g_4
=
\left(\begin{matrix}
1&0&0\\
0&\zeta_3&0\\
0&0&\zeta_3\\
\end{matrix}\right),
\end{align*}
where $\zeta_3$ denotes the primitive third root of 1.
The name, ``Hessian" group, comes from the fact that $G_{216}$ is the group of linear automorphisms acting on the Hesse pencil \cite{AD06,SS31}.
The Hesse pencil is a one-dimensional linear system of plane cubic curves in $\P^2(\C)$ given by 
\begin{align*}
f(x_0,x_1,x_2; t_0,t_1)
:=
t_0\left(x_0^3+x_1^3+x_2^3\right)+t_1 x_0x_1x_2
=
0,
\end{align*}
where $(x_0,x_1,x_2)$ is the homogeneous coordinate of $\P^2(\C)$ and the parameter $(t_0,t_1)$ ranges over $\P^1(\C)$ \cite{AD06}. 
The curve composing the pencil is called the Hesse cubic curve (see figure \ref{fig:HesseCurve}).

Each member of the pencil is denoted by $E_{t_0,t_1}$ and the pencil itself by $\left\{E_{t_0,t_1}\right\}_{(t_0,t_1)\in\P^1(\C)}$.
The nine  base points of the pencil are given as follows
\begin{align*}
&p_0=(0,1,-1)&
&p_1=(0,1,-\zeta_3)&
&p_2=(0,1,-\zeta_3^2)\\
&p_3=(1,0,-1)&
&p_4=(1,0,-\zeta_3^2)&
&p_5=(1,0,-\zeta_3)\\
&p_6=(1,-1,0)&
&p_7=(1,-\zeta_3,0)&
&p_8=(1,-\zeta_3^2,0).
\end{align*}
Any smooth curve in the pencil has the nine base points as its inflection points, and hence they are in the Hesse configuration 
\cite{AD06,SS31}.
We choose $p_0$ as the unit of addition of the points on the Hesse cubic curve.

The group $E_{t_0,t_1}[3]$ of three torsion points on $E_{t_0,t_1}$ consists of the nine base points $p_0,p_1,\cdots,p_8$.
The map
\begin{align*}
p_1 \longmapsto (1,0)
\qquad
p_3 \longmapsto (0,1)
\end{align*}
induces the group isomorphism $E_{t_0,t_1}[3]\simeq \Gamma$, which is the normal subgroup of $G_{216}$ generated by the elements $g_1$ and $g_2$.
Therefore, the action of $\Gamma$ fixes the parameter $(t_0,t_1)$ of the Hesse pencil.
Let $\alpha: G_{216}\to PGL(2,\C)$ be a map given by
\begin{align*}
\alpha(g):\ 
(t_0,t_1)
=
(x_0x_1x_2, x_0^3+x_1^3+x_2^3)
\longmapsto
(t_0^\prime,t_1^\prime)
=
(x_0^\prime x_1^\prime x_2^\prime , x_0^{\prime3}+x_1^{\prime3}+x_2^{\prime3}),
\end{align*}
where $g\in G_{216}$ and $g: (x_0, x_1, x_2)\mapsto(x_0^\prime, x_1^\prime, x_2^\prime)$.
Then we have ${\rm Ker}(\alpha)\supset\Gamma=\langle g_1,g_2\rangle$.
Actually, $\Gamma$ is a subgroup of ${\rm Ker}(\alpha)$ of index two.

On the other hand, $\alpha(g_3)$ and $\alpha(g_4)$ act effectively on $PGL(2,\C)$: 
\begin{align*}
&\alpha(g_3):\ 
(t_0,t_1)
\longmapsto
(t_0^\prime,t_1^\prime)
=
(3t_0+t_1,18t_0-3t_1)\\
&\alpha(g_4):\ 
(t_0,t_1)
\longmapsto
(t_0^\prime,t_1^\prime)
=
(t_0,\zeta_3^2t_1).
\end{align*}
Thus $g_3$ and $g_4$ induce the action on the Hesse pencil independent of its additive group structure.
We can easily check the following relation
\begin{align*}
\alpha(g_3)^2=\alpha(g_4)^3=1.
\end{align*}
It follows that we have
\begin{align*}
\alpha(G_{216})
=
\left\langle
\alpha(g_3),\alpha(g_4)
\right\rangle
\simeq
\mathcal{T},
\end{align*}
where $\mathcal{T}$ is the tetrahedral group.
Thus the group $\alpha(G_{216})$ acts on $PGL(2,\C)$ as the permutations among the following 12 elements
\begin{align*}
&\lambda:=\frac{t_1}{t_0},\
\zeta_3\lambda,\
\zeta_3^2\lambda,\
\frac{18-3\lambda}{3+\lambda},\
\frac{18\zeta_3-3\zeta_3\lambda}{3+\lambda},\
\frac{18\zeta_3^2-3\zeta_3^2\lambda}{3+\lambda},\
\frac{18-3\zeta_3\lambda}{3+\zeta_3\lambda},\\
&\frac{18\zeta_3-3\zeta_3^2\lambda}{3+\zeta_3\lambda},\
\frac{18\zeta_3^2-3\lambda}{3+\zeta_3\lambda},\
\frac{18-3\zeta_3^2\lambda}{3+\zeta_3^2\lambda},\
\frac{18\zeta_3-3\lambda}{3+\zeta_3^2\lambda},\
\frac{18\zeta_3^2-3\zeta_3\lambda}{3+\zeta_3^2\lambda}.
\end{align*}

The Hesse pencil contains four singular members with multiplicity three corresponding to the following $(t_0,t_1)$ \cite{AD06}
\begin{align*}
(t_0,t_1)
=
(0,1),\ 
(1,-3),\ 
(1,-3\zeta_3^2),\ 
(1,-3\zeta_3).
\end{align*}
Denote these points by $s_i$ ($i=1,2,3,4$) in order.
These $s_i$'s are permuted by $\alpha(G_{216})$ as follows
\begin{align}
&\alpha(g_3):\ 
s_1\longleftrightarrow s_2,
\quad
s_3\longleftrightarrow s_4
\label{eq:actalpg3}\\
&\alpha(g_4):\ 
s_2\longrightarrow s_3\longrightarrow s_4\longrightarrow s_2
\quad
\mbox{($s_1$ is fixed.)}
\label{eq:actalpg4}
\end{align}
Thus $s_i$'s can be corresponded to the vertices of the tetrahedron on which $\mathcal{T}\simeq \alpha(G_{216})$ acts. 

Moreover, let
\begin{align*}
g_0
=
\left(\begin{matrix}
1&0&0\\
0&0&1\\
0&1&0\\
\end{matrix}\right).
\end{align*}
Then we have
\begin{align*}
g_3^2
=
\left(g_4g_0\right)^3
=g_0.
\end{align*}
Therefore, we obtain
\begin{align*}
G_{216}/\Gamma
=
\left\langle
g_3,g_4
\right\rangle
\simeq
\tilde{\mathcal{T}},
\end{align*}
where $\tilde{\mathcal{T}}$ is the binary tetrahedral group.
Since $\tilde{\mathcal{T}}$ is isomorphic to $SL(2,\F_3)$, we obtain the semi-direct product decomposition $G_{216}\simeq \Gamma\rtimes SL(2,\F_3)$.

The level-three theta functions $\theta_0(z,\tau)$, $\theta_1(z,\tau)$, and $\theta_2(z,\tau)$ are defined by using the theta function $\vartheta_{(a,b)}(z,\tau)$ with characteristics:
\begin{align*}
\theta_{k}(z,\tau)
:=
\vartheta_{\left(\frac{k}{3}-\frac{1}{6},\frac{3}{2}\right)}(3z,3\tau)
=
\sum_{n\in\Z}e^{3\pi i\left(n+\frac{k}{3}-\frac{1}{6}\right)^{2}\tau}
e^{6\pi i\left(n+\frac{k}{3}-\frac{1}{6}\right)\left(z+\frac{1}{2}\right)}
\quad(k=0,1,2),
\end{align*}
where $z\in\C$ and $\tau\in\H:=\{\tau\in\C\ |\ {\rm Im}\mkern2mu \tau>0\}$.
Fixing $\tau\in\H$, we abbreviate $\theta_k(z,\tau)$ and $\theta_k(0,\tau)$ as $\theta_k(z)$ and $\theta_k$ for $k=0,1,2$, respectively.
We can easily see that the following holds
\begin{align}
&\theta_0=-\theta_1
\qquad\theta_2=0.
\label{eq:thetaprop4}
\end{align}

Let $L_\tau:=(-\tau)\Z+(3\tau+1)\Z$ be a lattice in $\C$.
Consider a map $\varphi:\C\to\P^2(\C)$,
\begin{align*}
\varphi:\ z\longmapsto (\theta_2(z),\theta_0(z),\theta_1(z)).
\end{align*}
This induces an isomorphism from the complex torus $\C/L_\tau$ to the Hesse cubic curve $E_{{\theta_2^\prime},6{\theta_0^\prime}}$.
It also induces the additive group structure on $E_{{\theta_2^\prime},6{\theta_0^\prime}}$ from $\C/L_\tau$ through the addition formulae for the level-three theta functions \cite{KKNT09};
let $(x_0,x_1,x_2)$ and $(x_0^\prime,x_1^\prime,x_2^\prime)$ be points on $E_{{\theta_2^\prime},6{\theta_0^\prime}}$, then the addition $(x_0,x_1,x_2)+(x_0^\prime,x_1^\prime,x_2^\prime)$ of the points is given as follows
\begin{align}
(x_0,x_1,x_2)+(x_0^\prime,x_1^\prime,x_2^\prime)
&=
(
x_1x_2{x_2^\prime}^2-x_0^2x_0^\prime x_1^\prime,
x_0x_1{x_1^\prime}^2-x_2^2x_0^\prime x_2^\prime,
x_0x_2{x_0^\prime}^2-x_1^2x_1^\prime x_2^\prime
).
\label{eq:addform}
\end{align}
The relation \eqref{eq:thetaprop4} implies that the unit of addition on $E_{{\theta_2^\prime},6{\theta_0^\prime}}$ induced by $\varphi$ is $p_0$:
\begin{align*}
\varphi:\ 0\longmapsto (\theta_2,\theta_0,\theta_1)=(0,1,-1)=p_0.
\end{align*}

By using \eqref{eq:addform}, we see that the actions of $g_1$ and $g_2$ on $E_{{\theta_2^\prime},6{\theta_0^\prime}}$ can be realized as the additions with $p_6$ and $p_1$, respectively
\begin{align*}
&(x_0,x_1,x_2)
\quad
\overset{g_1}{\longmapsto}
\quad
(x_1,x_2,x_0)
=
(x_0,x_1,x_2)+p_6\\
&(x_0,x_1,x_2)
\quad
\overset{g_2}{\longmapsto}
\quad
(x_0,\zeta_3 x_1,\zeta_3^2 x_2)
=
(x_0,x_1,x_2)+p_1.
\end{align*}

Take the following representatives $z_{0k},z_{k1},z_{k2}$ of the zeros of $\theta_k(z)$ in $\C/L_\tau$ for $k=0,1,2$
\begin{align*}
\left(
\begin{matrix}
z_{20}&z_{21}&z_{22}\\[5pt]
z_{00}&z_{01}&z_{02}\\[5pt]
z_{10}&z_{11}&z_{12}\\[5pt]
\end{matrix}
\right)
=
\left(
\begin{matrix}
0&\tau+\frac{1}{3}&2\tau+\frac{2}{3}\\[5pt]
-\frac{\tau}{3}&\frac{2\tau}{3}+\frac{1}{3}&\frac{5\tau}{3}+\frac{2}{3}\\[5pt]
-\frac{2\tau}{3}&\frac{\tau}{3}+\frac{1}{3}&\frac{4\tau}{3}+\frac{2}{3}\\[5pt]
\end{matrix}
\right).
\end{align*}
Then these nine zeros are mapped into the nine inflection points on $E_{{\theta_2^\prime},6{\theta_0^\prime}}$ by  $\varphi$, respectively:
\begin{align*}
\varphi:\quad
\begin{matrix}
z_{20}&z_{21}&z_{22}\\
z_{00}&z_{01}&z_{02}\\
z_{10}&z_{11}&z_{12}\\
\end{matrix}
\quad
\longmapsto
\quad
\begin{matrix}
p_0&p_1&p_2\\
p_3&p_4&p_5\\
p_6&p_7&p_8\\
\end{matrix}\ .
\end{align*}

Let us tropicalize the Hesse pencil.
For the defining polynomial $f(x_0,x_1,x_2; t_0,t_1)$ of the Hesse cubic curve, we apply the procedure of tropicalization.
Replacing $+$ and $\times$ with $\max$ and $+$ respectively, the polynomial $f(x_0,x_1,x_2; t_0,t_1)$ reduces to
\begin{align*}
\tilde f(\tilde x_0,\tilde x_1,\tilde x_2; \tilde t_0,\tilde t_1)
=
\max\left(
\tilde t_0+3\tilde x_0,\tilde t_0+3\tilde x_1,\tilde t_0+3\tilde x_2,\tilde t_1+\tilde x_0+\tilde x_1+\tilde x_2
\right).
\end{align*}
In order to distinguish tropical variables form original ones, we ornament them with $\tilde{}\ $.

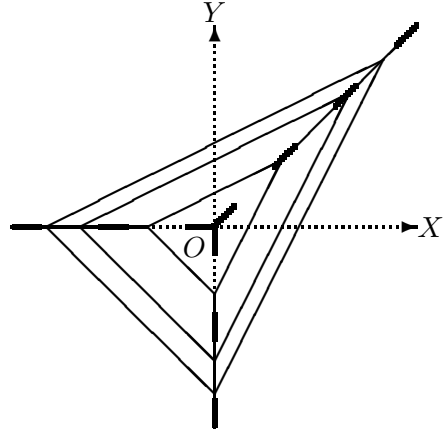
\begin{wrapfigure}[19]{r}{7cm}
\centering
{\unitlength=.035in{\def\arraystretch{1.0}
\begin{picture}(50,62)(-25,-30)
\thicklines
\put(0,28){\vector(0,1){2}}
\dottedline(0,-30)(0,28)
\put(0,32){\makebox(0,0){$Y$}}
\put(28,0){\vector(1,0){2}}
\dottedline(-30,0)(28,0)
\put(32,0){\makebox(0,0){$X$}}
\linethickness{1.6pt}
\dashline[10]{5}(0,0.1)(30,30.1)
\dashline[10]{5}(0,-0.1)(30,29.9)
\dashline[10]{5}(0,0)(0,-30)
\dashline[10]{5}(0,0)(-30,0)
\thicklines
\put(-25,0){\line(1,-1){25}}
\put(0,-25){\line(1,2){25}}
\put(-25,0){\line(2,1){50}}
\put(-20,0){\line(1,-1){20}}
\put(0,-20){\line(1,2){20}}
\put(-20,0){\line(2,1){40}}
\put(-10,0){\line(1,-1){10}}
\put(0,-10){\line(1,2){10}}
\put(-10,0){\line(2,1){20}}
\put(-10,0){\line(-1,0){20}}
\put(0,-10){\line(0,-1){20}}
\put(10,10){\line(1,1){20}}
\put(-3,-3){\makebox(0,0){$O$}}
\end{picture}
}}
\caption{
The curves drawn with solid lines are the regular members and the one with broken line is the singular member of the tropical Hesse pencil.
}
\label{fig:tropHesse}
\end{wrapfigure}

Let $(t_0,t_1)$ be a point in $\P^{1,trop}$, the tropical projective line.
Then $\tilde f$ can be regarded as a function $\tilde f:\P^{2,trop}\to\T$, where $\P^{2,trop}$ is the tropical projective plane and $\T:=\R\cup\{-\infty\}$ is the tropical semi-field.
The tropical Hesse curve is the set of points such that the function $\tilde f$ is not differentiable. 
We denote the tropical Hesse curve by $C_{\tilde{t}_0,\tilde{t}_1}$.
Upon introduction of the inhomogeneous coordinate $(X:=\tilde{x}_1-\tilde{x}_0,Y:=\tilde{x}_2-\tilde{x}_0)\in\P^{2,trop}$ and $K:=\tilde{t}_1-\tilde{t}_0\in\P^{1,trop}$ the tropical Hesse curve is denoted by $C_K$ and is given by the tropical polynomial
\begin{align*}
F(X,Y; K)
:=
\max\left(
3X,3Y,0,K+X+Y
\right).
\end{align*}

Figure \ref{fig:tropHesse} shows the tropical Hesse curves.
The one-dimensional linear system $\{C_K\}_{K\in\P^{1,trop}}$ consisting of the tropical Hesse curves is called the tropical Hesse pencil.
The complement of the tentacles, i.e., the finite part, of $C_K$ is denoted by $\bar C_K$. 
We denote the vertices whose coordinates are $(K,K)$, $(-K,0)$, and $(0,-K)$ by $V_1$, $V_2$, and $V_3$, respectively.

Let $K$ and $\varepsilon$ be positive numbers.
Let us fix $\tau$:
\begin{align*}
\tau
=
-\frac{3K}{9K+2\pi i \varepsilon}.
\end{align*}
Then the complex torus $\C/L_{\tau}$ converges into $J(C_K)$ in the limit $\varepsilon\to0$ with respect to the Hausdorff metric \cite{N11}.
Let us introduce a map $\tilde\varphi: J(C_K)\to \R^2\subset\P^{2,trop}$,
\begin{align*}
\tilde\varphi:\quad
u\longmapsto(\tilde c(u),\tilde s(u)),
\end{align*}
where we define
\begin{align*}
&\tilde c(u)
:=
-\frac{9K}{2}
\left\{
\left(\left(
\frac{u-K}{3K}-\frac{1}{2}
\right)\right)
\right\}^2
+
\frac{9K}{2}
\left\{
\left(\left(
\frac{u-3K}{3K}-\frac{1}{2}
\right)\right)
\right\}^2\\
&\tilde s(u)
:=
-\frac{9K}{2}
\left\{
\left(\left(
\frac{u-2K}{3K}-\frac{1}{2}
\right)\right)
\right\}^2
+
\frac{9K}{2}
\left\{
\left(\left(
\frac{u-3K}{3K}-\frac{1}{2}
\right)\right)
\right\}^2,
\end{align*}
and $\left(\left(u\right)\right):=u-{\rm Floor}(u)$.
This map induces an isomorphism $\bar C_K\simeq J(C_K)$ \cite{KKNT09,N11}, where $J(C_K)$ is the tropical Jacobian of $C_K$:
\begin{align*}
J(C_K)
:=
\R/3K\Z
=
\{u\in \R\ |\ 0\leq u<3K\}.
\end{align*}
Thus $\tilde\varphi$ induces additive group structure on $\bar C_K$ equipped with the unit of addition $V_1=\tilde\varphi(0)$ form $J(C_K)$.
The addition formula for $\bar C_K$ is explicitly given in \cite{N11}.

The piecewise linear functions $\tilde c(u)$ and $\tilde s(u)$ are periodic with period $3K$ and are the ultradiscretization of the elliptic functions $c(z):={\theta_{0}(z,\tau)}/{\theta_{2}(z,\tau)}$ and $s(z):={\theta_{1}(z,\tau)}/{\theta_{2}(z,\tau)}$, respectively \cite{KKNT09,N11}.
In the procedure of ultradiscretization, we assume $u\in\R$ and 
\begin{align*}
z
=
\frac{\left(1-i\xi_\varepsilon \right)u}
{9K},
\end{align*}
where $\xi_\varepsilon={2\pi\varepsilon}/{9K}$, and take the limit $\varepsilon\to0$. 
In terms of the variable $u$, we put the limit of zeros $z_{kj}$ $(k,j=0,1,2)$ of the level-three theta functions as follows
\begin{align*}
&u_2:=\lim_{\varepsilon\to0}9Kz_{20}=\lim_{\varepsilon\to0}9Kz_{21}=\lim_{\varepsilon\to0}9Kz_{22}=0
\\
&u_0:=\lim_{\varepsilon\to0}9Kz_{00}=\lim_{\varepsilon\to0}9Kz_{01}=\lim_{\varepsilon\to0}9Kz_{02}=K
\\
&u_1:=\lim_{\varepsilon\to0}9Kz_{10}=\lim_{\varepsilon\to0}9Kz_{11}=\lim_{\varepsilon\to0}9Kz_{12}=2K,
\end{align*}
where it should be noted that $\tau\to-1/3$ in the limit $\varepsilon\to0$. 

Consider a map $\eta:E_{{\theta_2^\prime},6{\theta_0^\prime}}\to \bar C_K$ so defined that the diagram commute
\begin{align*}
\begin{CD}
\C/L_\tau
@ > \varepsilon\to0 >> 
J(C_K)\\
@ V \varphi VV
@ VV \tilde\varphi V\\
E_{{\theta_2^\prime},6{\theta_0^\prime}}
@ > \eta >>
\bar C_K.\\
\end{CD}
\end{align*}
The inflection points of $E_{{\theta_2^\prime},6{\theta_0^\prime}}$ are mapped into the vertices of $\bar C_K$ by $\eta$ as follows
\begin{align}
&\eta:\ 
p_0,\ p_1,\ p_2
\overset{\varphi^{-1}}{\longmapsto}
z_{20},\ z_{21},\ z_{22}
\overset{\varepsilon\to0}{\longrightarrow}
u_2
\overset{\tilde\varphi}{\longmapsto}
V_1
\label{eq:eta1}\\
&\eta:\ 
p_3,\ p_4,\ p_5
\overset{\varphi^{-1}}{\longmapsto}
z_{00},\ z_{01},\ z_{02}
\overset{\varepsilon\to0}{\longrightarrow}
u_0
\overset{\tilde\varphi}{\longmapsto}
V_2
\label{eq:eta2}\\
&\eta:\ 
p_6,\ p_7,\ p_8
\overset{\varphi^{-1}}{\longmapsto}
z_{10},\ z_{11},\ z_{12}
\overset{\varepsilon\to0}{\longrightarrow}
u_1
\overset{\tilde\varphi}{\longmapsto}
V_3.
\label{eq:eta3}
\end{align}

Now we investigate the tropical counterpart of the Hessian group $G_{216}\simeq \Gamma\rtimes SL(2,\F_3)$.
At first we consider $\Gamma\simeq\left(\Z/3\Z\right)^2$.
Note that $\Gamma=\langle g_1,g_2\rangle$ and the actions of $g_1$ and $g_2$ on $E_{t_0,t_1}$ is realized as the additions with $p_6$ and $p_1$, respectively.
Moreover, the group generated by the additions with $p_6$ and $p_1$ is nothing but $E_{t_0,t_1}[3]$, the group of three torsion points on $E_{t_0,t_1}$.

The correspondence \eqref{eq:eta1}, \eqref{eq:eta2}, and \eqref{eq:eta3} in terms of $\eta$ tells us that the addition with $p_6$ corresponds to that with $V_3$ on $C_K$, while that with $p_1$ vanishes in the limit $\varepsilon\to0$.
 (Note that  $V_1$ is the unit of addition on $C_K$.)
Since the addition with $p_3$ (resp. $V_2$) is equivalent to that with $2p_6$ (resp. $2V_3$), the tropical analogue of $\Gamma$ consists of the addition with $V_3$. 
Actually, it is the group $C_K[3]=\left\langle V_3\right\rangle$ of three torsion points on $C_K$, which is isomorphic to $\Z/3\Z$.
We denote the tropical analogue of a group $G$ by $trop(G)$:
\begin{align*}
trop(\Gamma)
\simeq
\Z/3\Z.
\end{align*}
The addition with $V_3$ is explicitly computed as follows
\begin{align*}
(X,Y)\uplus V_3
=
(X,Y)\uplus (0,-K)
=
(Y-X,-X),
\end{align*}
where we denote the addition on $C_K$ by $\uplus$ and apply the addition formula \cite{N11}
\begin{align*}
(X,Y)\uplus (X^\prime,Y^\prime)
=
\left(
\max\left(
Y,2X+X^\prime+Y^\prime
\right)
-
\max\left(
X+2X^\prime,2Y+Y^\prime
\right),\right.\quad\\
\left.
\max\left(
X+Y+2Y^\prime,X^\prime
\right)
-
\max\left(
X+2X^\prime,2Y+Y^\prime
\right)
\right).
\end{align*}

The group $trop(\Gamma)$ can also be obtained by applying the procedure of ultradiscretization directly to $g_1$ and $g_2$.
Let us consider the inhomogeneous coordinate $(x:=x_1/x_0,y:=x_2/x_0)$ of $\P^2(\C)$.
Let $g_1:(x,y)\mapsto (x^\prime,y^\prime)$ and $g_2:(x,y)\mapsto (x^{\prime\prime},y^{\prime\prime})$.
Then we have
\begin{align*}
\left(\left|x^\prime\right|,\left|y^\prime\right|\right)
=
\left(
\frac{|y|}{|x|},\frac{1}{|x|}
\right)
\qquad
\left(\left|x^{\prime\prime}\right|,\left|y^{\prime\prime}\right|\right)
=
\left(
|x|,|y|
\right).
\end{align*}
Replacing $|x|$ and $|y|$ with $e^{X/\varepsilon}$ and $e^{Y/\varepsilon}$ respectively and taking the limit $\varepsilon\to0$, we obtain
\begin{align*}
(X,Y)
\overset{\tilde g_1}{\longmapsto}
(Y-X,-X)
=
(X,Y)\uplus V_3
\qquad
(X,Y)
\overset{\tilde g_2}{\longmapsto}
(X,Y),
\end{align*} 
where we denote the action on $\P^{2,trop}$ induced form $g_1$ and $g_2$ by $\tilde g_1$ and $\tilde g_2$, respectively.

Next we consider $\alpha(G_{216})\simeq{\mathcal{T}}$.
Remember that each singular member $E_{s_i}$ ($i=1,2,3,4$) in the Hesse pencil corresponds to the vertex of the tetrahedron on which $\mathcal{T}$ acts (see \eqref{eq:actalpg3} and \eqref{eq:actalpg4}).
The singular members of the tropical Hesse pencil are $C_{\infty}$ and $C_0$ which are the tropicalization of $E_{s_1}$ and $E_{s_i}$, ($i=2,3,4$), respectively \cite{N11}.
Thus the action of $\alpha(g_3)$, which permutes $s_1$ and $s_3$ with $s_2$ and $s_4$ respectively, must vanish; while the action of $\alpha(g_4)$, which fixes $s_1$ and permutes $s_2$, $s_3$, and $s_4$ cyclically, reduces to the action fixing both $C_0$ and $C_\infty$.
Therefore, we have
\begin{align*}
trop\left(\alpha(G_{216})\right)
\simeq
trop\left(\mathcal{T}\right)
\simeq
\left\langle 1\right\rangle.
\end{align*}
Thus the tropical analogue of the Hessian group fixes each member of the tropical Hesse pencil.

Furthermore, we consider the tropicalization of $g_0=g_3^2$.
We ultradiscretize $g_0$ directly as well as $g_1$ and $g_2$.
The action of $g_0$ on $\P^2(\C)$ is given as
\begin{align*}
\left(
x,y
\right)
\overset{g_0}{\longmapsto}
\left(
{y},{x}
\right)
\end{align*}
in the inhomogeneous coordinate.
It follows that we have
\begin{align*}
\left(
X,Y
\right)
\overset{\tilde g_0}{\longmapsto}
\left(
{Y},{X}
\right),
\end{align*}
where $(X,Y)$ is the inhomogeneous coordinate of $\P^{2,trop}$ and $\tilde g_0$ is the action on $\P^{2,trop}$ induced from that of $g_0$ by applying the procedure of ultradiscretization.
Thus we conclude that the tropical analogue of $\tilde{\mathcal{T}}\simeq\langle g_3,g_4\rangle=G_{216}/\Gamma$ is the group of order two generated by $\tilde g_0$:
\begin{align*}
trop\left(\tilde{\mathcal{T}}\right)
\simeq
\left\langle \tilde g_0\right\rangle
\simeq
\left\langle \left(\begin{matrix}2&0\\0&2\\\end{matrix}\right)\right\rangle
\subset SL(2,\F_3).
\end{align*}

We then obtain the following theorem concerning the tropical analogue of $G_{216}$.
\begin{theorem}\normalfont
The dihedral group $\mathcal{D}_3$ of degree three,
\begin{align*}
\mathcal{D}_3
=\left\langle \tilde g_0, \tilde g_1\right\rangle
\simeq
\left(\Z/3\Z\right)
\rtimes
\left\langle
\left(
\begin{matrix}
2&0\\
0&2\\
\end{matrix}
\right)
\right\rangle
\end{align*}
where $\tilde g_0, \tilde g_1\in PGL(3,\T)$ satisfy $\tilde g_0^2 =\tilde g_1^3=\left(\tilde g_0\tilde g_1\right)^2=1$, is the group of linear automorphisms acting on the tropical Hesse pencil\footnote{
Identifying $\gamma_i\simeq (\gamma_i,0)\in\left(\Z/3\Z\right)^2$, we define the multiplication of  $(\gamma_1,m_1), (\gamma_2,m_2)\in \left(\Z/3\Z\right)\rtimes \left\langle\left(\begin{matrix}2&0\\0&2\\ \end{matrix}\right)\right\rangle$ by 
\begin{align*}
(\gamma_1,m_1)\cdot(\gamma_2,m_2)
=
(\gamma_1+m_1\gamma_2, m_1m_2).
\end{align*}
}.
The action of $\tilde g_0$ on each curve of the pencil is realized as the reflection with respect to the line $Y=X$ passing through the vertex $V_1$; and the action of $\tilde g_1$ on each curve is realized as the addition with $V_3$.
\end{theorem}

In this paper, we consider linear automorphisms acting on the Hesse pencil only.
To investigate a tropical analogue of the Cremona group, the group of birational automorphisms acting on the Hesse pencil, is a further problem.


\end{document}